\documentclass[sigconf]{acmart}

\copyrightyear{2026}
\acmYear{2026}
\setcopyright{cc}
\setcctype{by}
\acmConference[ICSE '26]{2026 IEEE/ACM 48th International Conference on Software Engineering}{April 12--18, 2026}{Rio de Janeiro, Brazil}
\acmBooktitle{2026 IEEE/ACM 48th International Conference on Software Engineering (ICSE '26), April 12--18, 2026, Rio de Janeiro, Brazil}
\acmDOI{10.1145/3744916.3773188}
\acmISBN{979-8-4007-2025-3/2026/04}

\usepackage{csquotes}           
\usepackage{glossaries-extra}   
\usepackage{caption}            
\usepackage{dblfloatfix}        

\newcommand*{\resultsSectionHeader}[4]{%
    \begin{figure*}[t]
        \centering
        \includegraphics[width=1\linewidth]{#2}
        \Description{#4}
        \label{#3}
    \end{figure*}
    \subsection{#1}
}
\newcommand*{\jpq}[3]{\enquote{#1}~{\scriptsize[#2]}}
\newcommand*{\jpqid}[1]{{\scriptsize#1}}

\newcommand*{\varLinesOfText}{4,165}

\glsdisablehyper        

\setabbreviationstyle{long-short}
\newabbreviation{se}{SE}{Software Engineering}

\setabbreviationstyle[long]{long}
\newabbreviation[category=long]{ta}{}{Thematic Analysis}

\begin{document}

\title[What's in a Software Engineering Job Posting?]{What's in a Software Engineering Job Posting?}

\author{Marvin Wyrich}
\authornote{Both authors contributed equally to this research.}
\email{wyrich@cs.uni-saarland.de}
\orcid{0000-0001-8506-3294}
\affiliation{%
  \institution{Saarland University}
  \city{Saarbrücken}
  \country{Germany}
}
\author{{Lloyd} Montgomery}
\authornotemark[1]
\email{lloyd.montgomery@uni-hamburg.de}
\orcid{0000-0002-8249-1418}
\affiliation{%
  \institution{University of Hamburg}
  \city{Hamburg}
  \country{Germany}
}

\renewcommand{\shortauthors}{Wyrich and Montgomery}

\begin{abstract}
A well-rounded software engineer is often defined by technical prowess and the ability to deliver on complex projects.
However, the narrative around the ideal \gls{se} candidate is evolving, suggesting that there is more to the story.
This article explores the non-technical aspects emphasized in \gls{se} job postings, revealing the sociotechnical and organizational expectations of employers.
Our \gls{ta} of 100 job postings shows that employers seek candidates who align with their sense of purpose, fit within company culture, pursue personal and career growth, and excel in interpersonal interactions.
This study contributes to ongoing discussions in the \gls{se} community about the evolving role and workplace context of software engineers beyond technical skills.
By highlighting these expectations, we provide relevant insights for researchers, educators, practitioners, and recruiters.
Additionally, our analysis offers a valuable snapshot of \gls{se} job postings in 2023, providing a scientific record of prevailing trends and expectations.
\glsresetall  
\end{abstract}

\ccsdesc[300]{Software and its engineering~Programming teams}
\ccsdesc[300]{Human-centered computing~Empirical studies in collaborative and social computing}
\ccsdesc[300]{Social and professional topics~Project staffing}

\keywords{software engineering, job postings, non-technical competencies, employer expectations, hiring trends, recruitment}

\maketitle

\section{Introduction}
\gls{se} roles are evolving, as are the ways in which they are described to prospective candidates.
Today's job postings go well beyond technical requirements.
They communicate a broad spectrum of non-technical aspects such as collaboration, autonomy, growth opportunities, team culture, and organizational values~\cite{Rabelo:2022:NonTechSkillsSE}.
These descriptions reveal how companies position themselves, what they value in employees, and the workplace environment they foster.
While technical skills remain fundamental, it is increasingly clear that hiring decisions are shaped by a wider range of considerations.
In fact, educators and industry have long argued that technical expertise alone is not enough for a software engineer to succeed~\cite{Li:2015:WhatMakesAGreatSE,Rabelo:2022:NonTechSkillsSE}, with growing calls to prepare students for the interpersonal and cultural demands of software work~\cite{Garcia:2025:NonTechSkillsIndustryGap,Rabelo:2022:NonTechSkillsSE,Niva:2023:JuniorSECommCollab}.

These expectations are not hidden---they are already visible in plain sight, embedded in the language of everyday job postings.
Companies use such postings to attract candidates, but also to convey their identity and signal their values~\cite{Hein:2025:Branding}.
As such, job postings reveal more than hiring criteria, and offer a window into how companies envision the role of software engineers within their organizational and cultural context.
Looking through this window is valuable not only for jobseekers and educators, but also for researchers interested in how SE roles are framed in practice.

Prior research has begun to explore these trends, typically through quantitative methods such as keyword extraction, or frequency analysis~\cite{Ehlers:2015:Socialness,Montandon:2021:SkillsSOpostings}.
These studies have provided useful snapshots, especially of mentioned \emph{soft} and other \emph{non-technical skills}, but often rely on surface-level metrics, and offer limited interpretive depth or thematic structure.
While a few mixed-method studies have examined non-technical skills in SE hiring~\cite{Li:2015:WhatMakesAGreatSE,Rabelo:2022:NonTechSkillsSE}, much less attention has been paid to the more implicit, diverse, and conceptually rich non-technical content of job advertisements.

In this paper, we address that gap by conducting an inductive, qualitative analysis of 100 job postings from 100 companies worldwide.
Using reflexive \gls{ta}, we identify latent themes in non-technical aspects of SE job postings, including interpersonal expectations, support mechanisms, identity claims, and the broader sociotechnical context of SE work.
Our findings offer a nuanced account of what employers signal through job postings, what they promise, what they prioritize, and what they leave unsaid.

\section{Methodology}

We are interested in the non-technical aspects that employers highlight in \gls{se} job postings.
We sampled 100 job postings and applied \gls{ta} to answer this singular research question:

\begin{description}
    \item[RQ] What non-technical aspects are being discussed in \glsxtrlong{se} job postings?
\end{description}

\subsection{Sampling Job Postings}

We sampled 100 full-time \gls{se} job postings from five different job platforms: LinkedIn, SimplyHired, Monster, Indeed, and Glassdoor.
We collected 20 postings per platform from 100 distinct companies.
To minimize platform-specific biases, we searched for \enquote{Software Engineer} worldwide, sorted results by date, and selected the 20 most recent full-time positions from unique companies.
Searches were conducted in incognito mode from a German IP address, and only English-language postings were included.
For platforms with country-specific subdomains, we randomly distributed selections across supported countries.
Meta-data, full-text content, and PDFs were recorded for each posting, with website archiving~\cite{InternetArchive_2025} where possible.
Further details on the sampling procedure are provided in the replication package~\cite{ReplicationPackage}.
We provide a summary of the metadata in Section~\ref{sec:metadata}.

\subsection{Thematic Analysis}

To reveal the non-technical aspects described in \gls{se} job postings, the first two authors conducted an \textit{inductive} \gls{ta} of the job postings, producing a \textit{rich overview} with \textit{latent} themes.
We followed the recommendations of Braun and Clarke~\cite{Braun:2006:thematic}.
We first decided on three core factors that guided the \gls{ta} process: level of detail, style of approach, and level of analysis~\cite{Braun:2006:thematic}.
We decided on a ``rich overview'' level of detail to capture the entire landscape of non-technical aspects being discussed, rather than focusing on one specific sub-area.
We chose an ``inductive'' style of approach to identify themes that are strongly linked to the data, and not driven by our own theoretical or analytical interest in the area.
We chose the ``latent'' level of analysis as we were interested in going beyond the semantic ``surface meanings of the data'' and instead identifying ``the underlying ideas, assumptions, and conceptualizations [that inform] the semantic content of the data''~\cite{Braun:2006:thematic}.

In \textbf{phase one} (of the five phases of \gls{ta}~\cite{Braun:2006:thematic}), we familiarised ourselves with the {\varLinesOfText} lines of text extracted from the job postings---which we call evidence---by reading each line carefully.
With so much evidence, it was important to read and digest the full extent of information in the dataset, before taking notes or assigning initial codes.
Both researchers conducted this phase independently, taking 10 hours each, before meeting to discuss their observations.
In \textbf{phase two}, we generated an initial set of codes in chunks of ten job postings at a time, starting with re-reading and taking notes on each line regarding potential codes, followed by another full read to assign codes.
We did this two-read pass to re-familiarise ourselves with the evidence, and to mitigate the learning effect during open labelling.
We used the Framework Method~\cite{Gale_2013_BMCMRM} to record and link evidence for each code, such that full evidence tracing through our codes, sub-themes, and themes is possible via our replication package~\cite{ReplicationPackage}.
Both researchers conducted this phase independently, taking 27 hours each.
In \textbf{phase three}, we created the initial themes first independently, followed by shared sessions to merge everything into an initial thematic map.
The result of this phase was a set of themes, sub-themes, and codes.
The codes were detailed enough, and the themes were high-level enough, that an intermediate layer of sub-themes was needed to provide a connecting structure between them.
This phase took 5 hours each.

In \textbf{phase four}, we iteratively reviewed and refined the themes on two levels: first on a per-theme level and then across all themes.
For level one, we investigated each theme separately, analysing their sub-themes and codes.
Our goal was to understand and refine all constructs, such that their definitions were descriptive and concise, while also not fracturing the themes into too many sub-concepts.
When a sub-theme or code did not fit well into the hierarchy, we re-visited the individual evidence pieces to re-familiarise ourselves with the data, and then we re-named, re-defined, and re-organised the concepts.
At the end of this phase, we knew that each theme was internally well-structured, and all evidence was appropriately placed.
This level one analysis took a long time due to the process of re-reading evidence pieces one at a time, while re-imagining the internal structural organisation of each theme.
This analysis took 73 hours of collaborative meetings (146 person-hours).
For level two, we investigated the themes in relation to each other, analysing the sub-themes and codes in relation to concepts presented across other themes.
Our goal was to decouple similar ideas across themes, creating more specific and unique ideas grouped under each theme.
This level of analysis also involved reviewing individual evidence pieces again, but much less so.
Level two took 15 hours of collaborative meetings.
In \textbf{phase five}, we revisited the names and definitions of our themes, sub-themes, and codes, looking to bring clarity and understandability to the concepts.
Given the in-depth and detailed nature of our analyses in phase four, there was not much work to be done here.
Phase five took 11 hours of collaborative meetings.

In total, the \gls{ta} took 282 person-hours.
The final 198 hours (70\%) was spent collaboratively analysing the evidence and iterating over the codes, themes, and thematic map.
Intercoder reliability in such exploratory and collaborative analysis is not meaningful, as initial alignment is not the goal.
Rather, the reflexive nature of the method embraces diverse researcher perspectives, which enrich the initial phases of \gls{ta}~\cite{McDonald_2019_CSCW,OConnor_2020_IJQM}.

\begin{figure*}[b]
    \centering
    \includegraphics[clip, trim={0 0 0 0.1cm}, width=1\linewidth]{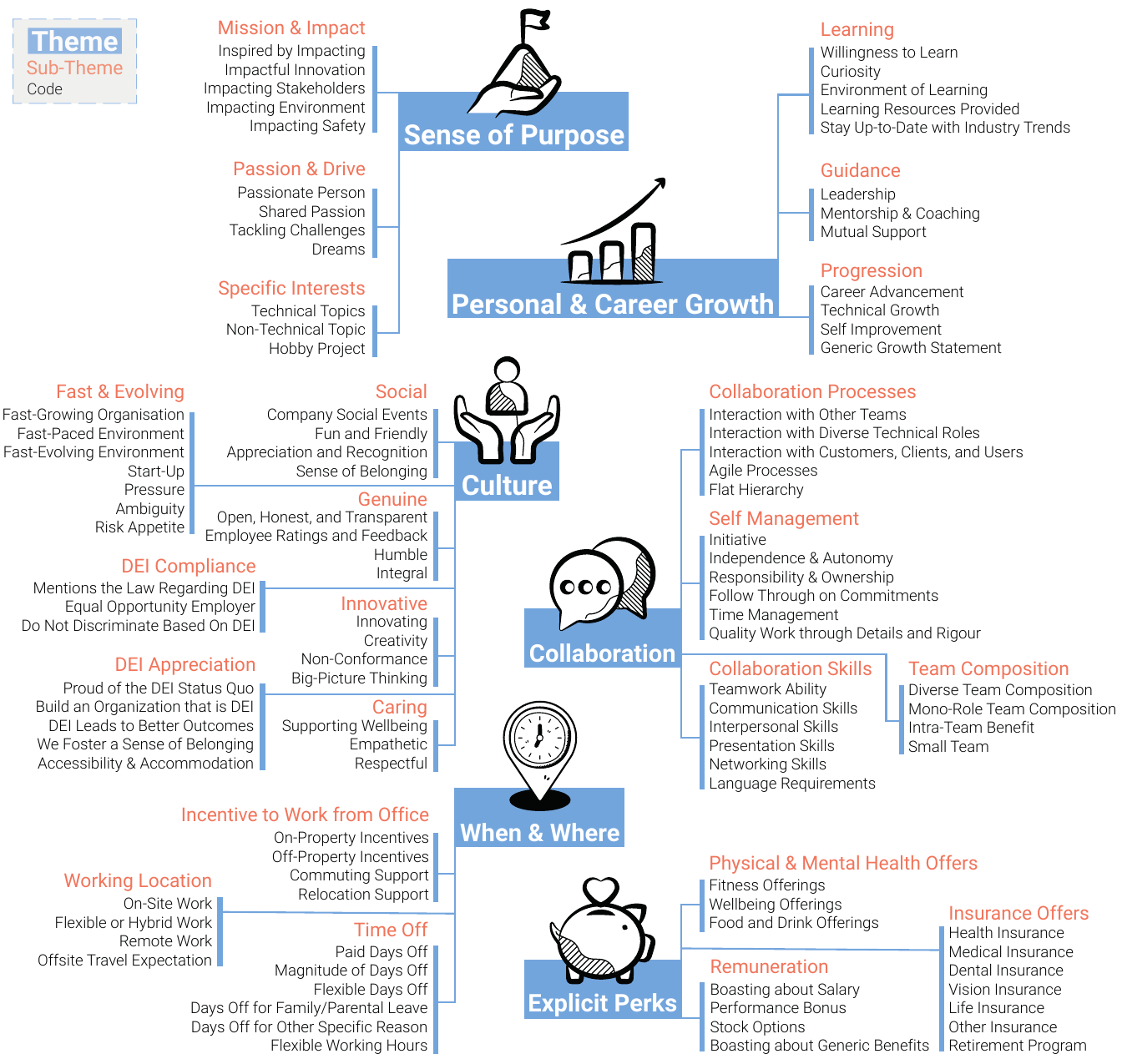}
    \caption{Thematic map of non-technical aspects in \glsxtrlong{se} job postings.}
    \Description{This figure displays a thematic map of all non-technical aspects we extracted from 100 Software Engineering job postings. The figure prominently displays six themes, each with a representative icon and a name. These six themes are Sense of Purpose, Personal \& Career Growth, Culture, Collaboration, When \& Where, and Explicit Perks. The figure then breaks these six themes down into their sub-themes, of which there are 23. These 23 sub-themes are then broken down into their codes, of which there are 103.}
    \label{fig:thematic_map}
\end{figure*}

\subsection{Limitations}

\paragraph{Sampling Job Postings.} Sorting by date helps reduce platform bias, but ranking algorithms may still influence which postings appear.
Additionally, limiting the sample to English-language postings excludes local markets, potentially biasing regional representation.
Finally, while the sample may seem small, it aligns with the aims of reflexive \gls{ta}, which prioritizes conceptual richness over saturation~\cite{Braun:2021:Commentary}.
A sample of 100 job postings (with {\varLinesOfText} lines of evidence) supports such richness while keeping the scope manageable for our desired in-depth, interpretive engagement.

\paragraph{Thematic Analysis.}
As with all forms of content analysis, our interpretations are shaped by our perspectives and may not capture the full range of possible meanings within the data.
Moreover, since our analysis relies solely on job postings, employer intentions behind specific phrasing choices remain unverified.
Finally, our findings represent a snapshot in time and may not fully capture future shifts in hiring practices, although they offer a valuable record in time for understanding these shifts, contributing to broader analyses of how employer expectations change over time.

\section{Results}

Figure~\ref{fig:thematic_map} depicts the resulting thematic map derived from our \gls{ta}.
The evidence in job postings resulted in 103 codes, which we combined into 23 sub-themes and six themes. 
Section~\ref{sec:metadata} provides an overview of the dataset.
Each theme is then examined in detail (Sections~\ref{sec:purpose}--\ref{sec:explicit_perks}) before concluding with cross-theme observations in Section~\ref{sec:cross-theme}.

\subsection{Metadata of 100 SE Job Postings}
\label{sec:metadata}

Our dataset consists of 100 job postings from 100 distinct companies, covering a rather homogeneous range of \gls{se} roles. The majority of job titles (58\%) explicitly mention \enquote{Software Engineer}, often with further specialization such as \enquote{Fullstack}, \enquote{Junior}, or \enquote{Python}. Another 16\% include \enquote{Developer} in some form, and an additional 8\% use closely related terms like \enquote{Software Development Engineer} or \enquote{System Engineer}. In total, 82\% of job postings in our sample describe general \gls{se} positions, while the remaining 18\% focus specifically on software testing and quality assurance, Dev(Sec)Ops, as well as reliability engineering and IT infrastructure support.
The list of companies includes big tech companies like Google and Microsoft, as well as medium-sized and smaller companies.

Geographically, the job postings span 39 countries across all continents, with the highest representation from the United States (13), Canada (7), United Kingdom (6), Belgium (5), and Germany (5).
Europe leads with 44 postings across 15 countries, followed by North America with 24 postings in four.
While this global coverage offers insights into worldwide \gls{se} hiring trends, the small per-region sample sizes limit meaningful comparisons between regions.

Job posting lengths vary significantly.
The shortest posting contains only 39 words, while the longest extends to 1,289 words.
The median posting length is 490 words, consisting of 31 lines of text.

\resultsSectionHeader{Sense of Purpose}{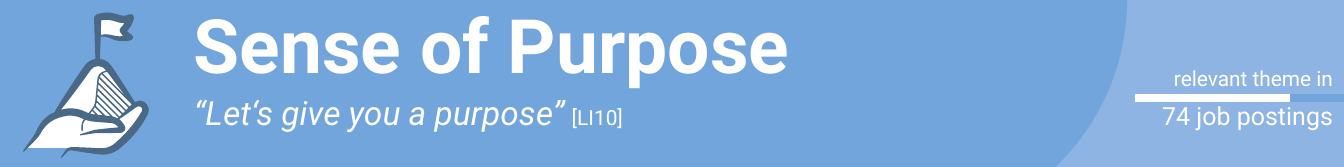}{fig:header_purpose}{Banner image indicating the start of the Sense of Purpose theme section. It features an icon representing the theme, the theme name in bold text, and the representative quote 'Let us give you a purpose' from job posting LI10. It indicates that 74 of the 100 analysed job postings include evidence related to this theme.}
\label{sec:purpose}

We begin our deep dive into \gls{se} job postings with a theme that often takes centre stage in the opening lines: \emph{Sense of Purpose}. This theme encompasses three sub-themes and 12 distinct codes. Evidence supporting these codes was identified in 74 out of the 100 job postings analysed.

\subsubsection{Mission \& Impact}

Are you \jpq{keen to see your work have an impact---fast?}{IN02}{T035}, or do you \jpq{feel satisfaction seeing what you've developed running in the real world?}{GL11}{T001}.
If so, you are the target audience for job postings that emphasize the importance of Mission \& Impact.
Many companies recognize the influence they have and highlight it in their job postings, often framing it as an invitation to prospective candidates.
This emphasis appears in various forms: a clear and specific mission, such as \jpq{our mission is to inspire creativity and bring joy}{SI05}{T048}; a more abstract drive, like \jpq{When you join us, you'll also be joining [our] global organization, where 80,000 people wake up every day determined to help our customers win}{GL07}{T017}; or a focus on high-impact projects where candidates help drive success, as in \jpq{Do you like working on industry-defining projects that move the needle?}{SI03}{T007}.

Codes within this sub-theme largely revolve around the potential and actual impact of the company in areas such as innovation, stakeholders, environment, and safety. These statements are designed to cast the hiring company in a powerful and influential light. The exception to this pattern is the code \emph{inspired by impacting}, which, while still emphasizing the significance of impact, places greater focus on the candidate. It highlights that this is the right place for individuals who are driven by the desire to make a difference (e.g.
\jpq{Now, imagine that you're the one who helps bring these constellations to life}{LI07}{T002}, \jpq{It's fun to work in a company where people truly BELIEVE in what they're doing!}{LI06}{T001}). This emphasis on purposefulness is taken to an extreme in one instance, where a job posting for a frontend \gls{se} position explicitly states, \jpq{Let's give you a purpose}{LI10}{T009}.

\subsubsection{Passion \& Drive}

Beyond the desire to make an impact, job postings emphasize further facets of personal drive and passion.
Passion is framed as a defining trait---whether through shared enthusiasm within teams, a culture of ambitious dreaming, or the desire to tackle tough challenges.
In this context, job postings frequently highlight that candidates will be surrounded by \jpq{like-minded professionals with great expertise and passion for what they do}{GL10}{T010} and \jpq{people from around the world that share the same passion and dedication}{LI16}{T040}, emphasizing that the candidate will not be alone in their enthusiasm.

This shared passion is often elevated beyond individual mo\-tivation---it becomes part of the company's culture. Some job postings explicitly cultivate an environment where ambitious dreaming is not just encouraged but expected: \jpq{You'll belong to a culture of dreamers, team players and avid learners with a flexible, value-based approach}{LI11}{T003}. Others frame big aspirations as a collective effort, stating, \jpq{We dream big together, supporting each other to make our individual and collective dreams come true}{MO10}{T006}. Even the hiring criteria can reflect this mindset, as one posting asserts, \jpq{We believe skills and experience are transferable, and the desire to dream big makes for great candidates}{MO10}{T014}.
In some cases, job postings invite candidates to envision the future alongside them: \jpq{Do you dream of what cars of the future will look like when you combine them with connectivity, a smartphone, and cloud services?}{IN08}{T002}, followed by the promise of \jpq{uniting those dreams with a company that has the skills and relationships to make that a reality}{IN08}{T003}.

\subsubsection{Specific Interests}

A few job postings go beyond broad notions of purpose and personal drive, highlighting specific interests that candidates should bring to the role. These range from enthusiasm for particular technical topics to engagement in non-technical domains and personal hobby projects.

Some postings signal that deep technical curiosity is valued, citing a \jpq{passion for engineering excellence through automation and process improvements}{LI13}{T049} or, more specifically, a \jpq{passion for performance debugging and benchmarking} {SI06}{T030}.
However, personal interests outside pure technical skills also matter.
Particularly in industries like gaming, job postings seek candidates who resonate with the field on a personal level, stating, \jpq{A love for gaming: You're a gamer at heart}{LI07}{T024} or even inviting candidates to \jpq{Let us know what your favorite game is}{LI05}{T043}.
Beyond gaming, postings highlight broader interests, such as \jpq{a general interest in books, reading and always learning new things}{LI11}{T041}.
Finally, personal projects, especially in open-source communities, appear as a recognized indicator of passion and dedication, with companies looking for \jpq{Contributions to open-source projects or a GitHub portfolio showcasing your hobby projects}{LI05}{T032}.

\subsubsection{Discussion}

The strong emphasis on sense of purpose in job postings reflects a broader trend in which many professionals seek meaningful and impactful work that extends beyond financial compensation~\cite{Beecham:2008:MotivationSE}.
Companies strategically position themselves as mission-driven organizations, leveraging purpose as both a branding tool and a means of attracting intrinsically motivated candidates.
However, this framing also carries implications for jobseekers.
While some may genuinely resonate with a company's mission, others might feel pressured to demonstrate passion or purpose alignment---even if their primary motivation lies in financial compensation, career development, or stability.

\resultsSectionHeader{Personal \& Career Growth}{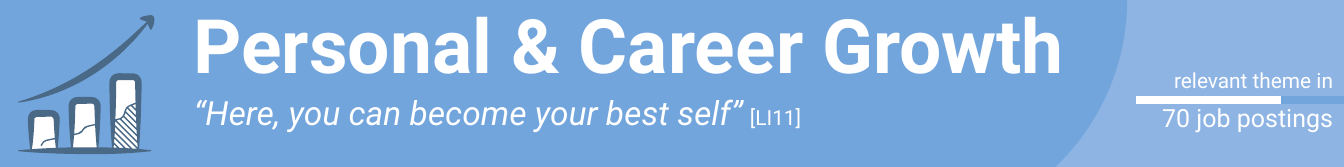}{fig:header_growth}{Banner image indicating the start of the Personal and Career Growth theme section. It features an icon representing the theme, the theme name in bold text, and the representative quote 'Here, you can become your best self' from job posting LI11. It indicates that 70 of the 100 analysed job postings include evidence related to this theme.}

The \textit{Personal \& Career Growth} theme encompasses three sub-themes and 12 codes.
We identified evidence supporting these codes in 70 out of the 100 job postings analysed.

\subsubsection{Learning}

Do you have an \jpq{appetite for learning}{LI13}{T036} and the \jpq{motivation to develop and enhance your skills}{LI20}{T019}?
If so, these job postings may entice you to apply.
\textit{Learning} is an important part of the \textit{Personal \& Career Growth} theme, focusing on the candidate as a life-long learner.
The Learning sub-theme has codes split between focusing on the individual's desire to learn and what the workplace will provide to support this learning.
On the one hand, the codes \textit{Willingness to Learn} and \textit{Curiosity} are specifically focused on what learning capabilities the candidate is bringing to the role.
For example, \jpq{eager to learn and grow your skill levels}{MO04}{T025} and \jpq{curious nature and desire to immerse in a complex engineering role}{IN10}{T025}.
On the other hand, learning was also expressed as an offer from the workplace, with the codes \textit{Environment of Learning} and \textit{Learning Resources Provided}.
These codes include evidence such as \jpq{[we] believe in fostering a culture of continuous learning and growth}{MO11}{T005} and \jpq{access to tuition reimbursement and learning and development resources}{SI04}{T054}.
Finally, the code \textit{Stay Up-to-Date with Industry Trends} expressed a mix of candidate-centric and workplace-centric evidence.
For example, some job postings requested that the candidate have \jpq{basic knowledge of industry-wide technology trends and best practices}{LI19}{T019}, while other workplaces offered \jpq{an opportunity to challenge yourself in exciting projects using the latest technology}{MO04}{T032}.

\subsubsection{Guidance}

If you thrive in roles where you \jpq{serve as a functional lead or technical owner}{MO10}{T031}, \jpq{mentor and coach fellow team members}{LI10}{T014}, or \jpq{listen to feedback and also provide supportive feedback to help others grow/improve}{SI15}{T031}, these job postings may align with your strengths.
\textit{Guidance} is a sub-theme that focuses on the relationship structures between people that support \textit{Personal \& Career Growth}.
The codes within this sub-theme are \textit{Leadership}, \textit{Mentorship \& Coaching}, and \textit{Mutual Support}.
Evidence for the Leadership code is managerial in nature, focusing on providing direction and seeking success.
For example, \jpq{we train and enable all our leaders to support their team towards achieving goals}{SI13}{T050} and \jpq{an ability to consistently harness targeted value from your team}{IN04}{T059}.
The evidence tells a story of contrived concepts such as \jpq{performance and integration of your team}{GL10}{T018} and \jpq{harness targeted value from your team}{IN04}{T059}.
This is somewhat shared by the \textit{Mentorship \& Coaching} code, which shifts from leadership to mentoring.
Examples include, \jpq{able to mentor and coach others in agile development}{SI18}{T018} and \jpq{accompanying and coach a team of developers}{GL15}{T025}.
The tone warms up as we shift to the \textit{Mutual Support} code, with evidence that describes helping and supporting each other.
For example, \jpq{we celebrate and support one another---from big and small things in life to big career moments}{SI09}{T089}.

\subsubsection{Progression}

Many job postings also highlight personal and career growth through qualities like
\jpq{a strong conviction to build long-term careers}{LI02}{T013}, a \jpq{desire to perform and grow as an engineer}{SI08}{T029}, and \jpq{having opportunities to grow and develop on the job}{GL11}{T030}.
These aspects align with the sub-theme of 
\textit{Progression}, which captures the forward movement in a candidate's career, both technically and personally.
This sub-theme has four codes, \textit{Career Advancement}, \textit{Technical Growth}, \textit{Self Improvement}, and \textit{Generic Growth Statement}.
Career Advancement evidence discusses the well-known concept of career aspirations.
Examples include postings looking for an \jpq{energetic and enthusiastic individual keen to develop a career in \gls{se}}{LI13}{T035} and offering \jpq{ample opportunity for career advancement}{IN14}{T034}.
The codes Technical Growth and Self Improvement, furthermore, discuss specific types of advancement.
For example, evidence for Technical Growth include offers for \jpq{professional development support to master your craft}{LI10}{T028} and desires for candidates that have a \jpq{clear ambition to master [their] chosen area of work}{MO13}{T044}.
Self Improvement evidence, by contrast, focuses on candidates who have \jpq{always wanted to beat [their] self best}{LI02}{T003} and \jpq{actively seeks ways to grow and be challenged using [\ldots] informal development channels}{LI03}{T010}.
Finally, the Generic Growth Statements code contains evidence discussing ``growth'', but without specific direction.
For example, \jpq{personal professional learning and development support}{GL18}{T037} and \jpq{extraordinary people need opportunities to grow}{MO15}{T063}.

\subsubsection{Discussion}

Our findings suggest that job postings emphasize not only what employers offer, but also what they expect from candidates. 
Across the three sub-themes, there is a clear mutual investment in career development: candidates should seek growth, while employers provide opportunities to facilitate it.

While technical advancement remains key, our findings underscore the equally important role of soft skills in career growth. 
Many job postings emphasize mentorship, leadership, and mutual support as essential to development. 
Growth is not just an individual pursuit but is nurtured through structured guidance and peer relationships. 
The presence of these elements in 70 out of 100 postings suggests that companies value engineers who not only refine their own skills but also help others to grow.
This raises questions about evolving expectations in \gls{se} careers.
As \gls{se} continues to demand lifelong learning, career progression may become more tied to interpersonal skills and the ability to uplift teams.
Future research could explore whether companies actively foster these dynamics in practice, or if they remain aspirational hiring ideals.

\resultsSectionHeader{Culture}{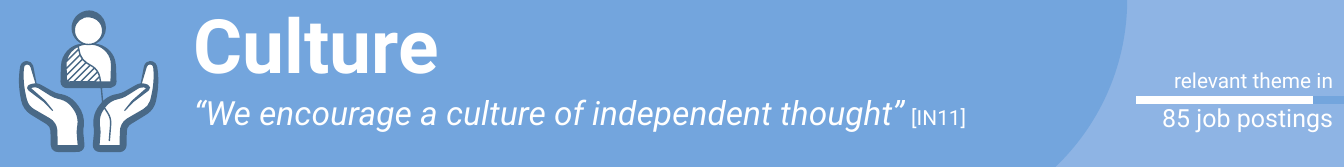}{fig:header_culture}{Banner image indicating the start of the Culture theme section. It features an icon representing the theme, the theme name in bold text, and the representative quote 'We encourage a culture of independent thought' from job posting IN11. It indicates that 85 of the 100 analysed job postings include evidence related to this theme.}

Company culture, defined as \enquote{the set of shared attitudes, values, goals, and practices that characterizes an institution or organization}~\cite{mw:culture}, shapes both employee experience and organizational identity. It is therefore no surprise that 85 out of 100 job postings we analysed highlighted aspects of their culture. Our analysis identified seven sub-themes, comprising a total of 30 distinct codes.

\subsubsection{Fast \& Evolving}
Given the rapid evolution of software technology, it is only natural that the software industry presents itself in job postings as fast-paced and fast-evolving. Some see this as an opportunity for inspiration (\jpq{You get inspired by and in a fast-paced environment where every day is a new day with new challenges faced.}{LI02}{T004}) or emerging possibilities (\jpq{Since we are increasing in size significantly, new opportunities are emerging constantly}{MO13}{T012}). 
However, this dynamic environment also translates into explicit requirements, such as \jpq{situational adaptability}{LI13}{T052}, the ability to work in a  \jpq{constantly changing environment}{SI14}{T022}, or prior experience \jpq{preferably in a fast-paced environment such as a startup or rapidly growing company}{LI10}{T017}.

Startups are often seen as embodying distinctive cultural aspects. While rapid growth is frequently mentioned, 
a noteworthy aspect is the potential for employees to shape the company culture: \jpq{As one of our early team members, you'll help to shape our company culture}{IN02}{T013}. A key trait in such settings is risk appetite---a mindset that embraces \jpq{taking calculated risks and embracing ambiguity as it comes}{SI05}{T008} and aligns with a \jpq{culture that inspires innovation, rewards risk-taking, and celebrates success}{SI15}{T007}.

High speed often brings high pressure.
Some job postings therefore appeal to candidates' resilience, requiring a \jpq{proven ability to work under pressure}{LI14}{T019}, being \enquote{comfortable with rapid change,} and showing a \jpq{willing[ness] to take ownership and deliver results in a high-pressure [...] environment}{IN03}{T064}.
Rarely, companies mention efforts to mitigate this pressure, such as \jpq{no customer deadlines---time to focus on quality, refactoring, and testing}{MO02}{T024}.

\subsubsection{Innovative}
Job postings often emphasize that innovative individuals and teams are part of the company and that an innovative culture prevails.
More specifically, this is framed in terms of creative thinking \jpqid{(e.g.~[SI08], [SI09], [MO04], [GL11])} and thinking outside the box \jpqid{(e.g.~[IN16], [GL19], [GL20])}, combined with big-picture thinking, as in, for example, the ability \jpq{to make trade-offs and technical decisions with a bigger roadmap in mind}{GL18}{T018}.
In rare cases, a degree of non-conformance is encouraged, with references to \jpq{a culture of independent thought, innovation, and problem-solving}{IN11}{T007}, an invitation to \jpq{Question and challenge---that's how we grow}{GL17}{T055}, or the guiding principle: \jpq{Dare to be different}{IN10}{T033}.

\subsubsection{Social, Genuine, and Caring}
Company culture is not always about speed or innovation.
The three sub-themes Social, Genuine, and Caring---supported by evidence across 11 distinct codes---focus more on interpersonal aspects that foster positive relationships and a supportive workplace atmosphere.

The \emph{Social} sub-theme includes statements about company social events and a fun and friendly working environment (e.g.,~\jpq{We organise after works and other team activities both online and offline and want to keep our friendly and informal atmosphere}{MO13}{}), as well as appreciation, recognition of good work, and a culture where employees feel a sense of belonging (e.g.,~\jpq{Join [our] close-knit team of tech enthusiasts, where your contributions will be recognized and rewarded}{MO11}{T008}).
\emph{Genuine} describes an open, honest, humble, and integrity-driven workplace, often supported by references to employee ratings and feedback (\jpq{We've achieved a lot but you'll find no airs and graces here}{GL17}{T046}).
Finally, \emph{Caring} highlights the importance of well-being and fostering a respectful, empathetic environment by prioritizing attention to others (\jpq{The well-being of our employees is a priority in our working culture!}{GL20}{T045}).

\subsubsection{DEI Compliance and Appreciation}
Our analysis revealed two cultural sub-themes related to diversity, equity, and inclusion (DEI): \emph{DEI Compliance} and \emph{DEI Appreciation}.
The first sub-theme generally involves a reference to a DEI law, a statement of being an \enquote{equal opportunity employer,} or a declaration of non-discrimination in hiring.
In some countries, such statements are legally required, explaining their presence in about a third of our sample.

However, we also found numerous evidence pieces that went beyond such formal compliance, reflecting a more genuine form of DEI appreciation.
Some companies express visible pride in their values, such as: \jpq{we don't just accept differences---we embrace them}{IN05}{T027}.
In some cases, this commitment is supported by data (e.g., \jpq{We are proud [...] that we have crushed our gender pay gap to 0.5\% in 2023}{IN10}{T042}, and \jpq{We truly live and celebrate our cultural diversity: our colleagues come from more than 70 countries and speak more than 40 languages}{GL08}{T035}).
Others stress that DEI is not merely for show but essential for better outcomes, \jpq{because without diversity and a dedication to equality, there is no moving forward}{SI09}{T092}.
This also means embracing individuality, as seen in 
\jpq{We love to see your uniqueness shine through and inspire the future of travel}{IN16}{T077}.

\subsubsection{Discussion}

A fast-paced, innovation-driven environment may fuel progress but can also clash with caring and inclusive aspects that emphasize well-being and belonging.
High adaptability and risk-taking expectations might drive innovation yet add pressure, raising the question of how companies reconcile these demands.
Some address this balance through perks, explored in Section~\ref{sec:explicit_perks}.
Meanwhile, external factors also shape how culture is communicated. 
Given current U.S. political trends, explicit DEI statements may become less common in postings from that region.

\resultsSectionHeader{Collaboration}{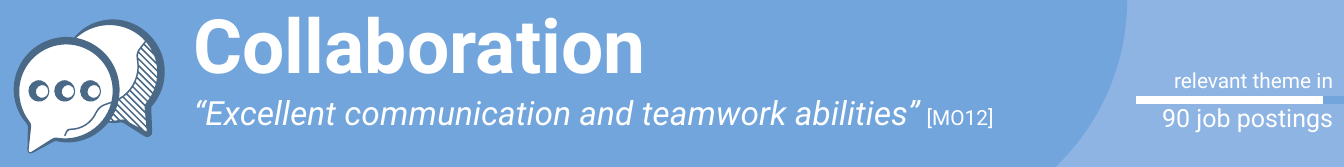}{fig:header_collaboration}{Banner image indicating the start of the Collaboration theme section. It features an icon representing the theme, the theme name in bold text, and the representative quote 'Excellent communication and teamwork abilities' from job posting MO12. It indicates that 90 of the 100 analysed job postings include evidence related to this theme.}

The \textit{Collaboration} theme encompasses four sub-themes and 21 codes.
We identified evidence supporting these codes in 90 of the 100 \gls{se} job postings analysed.

\subsubsection{Collaboration Processes}

Many job postings describe not just the need for collaboration, but the specific ways in which employees are expected to work together. Some candidates, for example, should be ready to \jpq{collaborate with cross-disciplinary teams to deliver successful solutions,}{LI06}{T007} and \jpq{collaborate with customers to find the perfect solution to their [needs]}{MO04}{T005}.
The \textit{Collaboration Processes} sub-theme encompasses evidence that describes the way in which people work together at their workplaces.
This sub-theme includes the codes \textit{Interaction with Other Teams}, \textit{Interaction with Diverse Technical Roles}, \textit{Interaction with Customers, Clients, and Users}, \textit{Agile Processes}, and \textit{Flat Hierarchy}.
Postings described \jpq{cross team collaboration with product and engineering teams}{LI06}{T007}, and the candidate's \jpq{ability to work collaboratively across teams}{MO07}{T021}.
They also described diverse technical roles that candidates need to interact with, including \jpq{product managers, data scientists, product engineers, and business leads}{SI08}{T022} and \jpq{QAs and UX designers}{MO07}{T017}.
This focus on stakeholder collaboration also extends to customers, clients, and users, for example, \jpq{direct interaction with [\ldots] open-source users and developer communities}{LI08}{T038}.
Job postings also mentioned Agile processes, although in the context of collaboration, not software development methodologies.
For example, \jpq{collaborates with other team members in agile processes}{LI03}{T012} and \jpq{we are a self-sufficient team that operates in a fully agile way}{GL14}{T020}.
Finally, some postings mentioned that they \jpq{don't have managers, no hierarchy}{SI17}{T028}, emphasising \jpq{a collegial work environment}{MO12}{T035} and \jpq{no boring manager... just great developments}{MO02}{T025}.

\subsubsection{Team Composition}

Do you want to \jpq{find yourself in a cross-functional team}{LI16}{T010}, or would you rather work in a specialised team such as \jpq{cloud experts}{GL02}{T003}, \jpq{core microservices}{GL03}{T012}, or \jpq{Multi-Cultural Compliance}{GL04}{T010}?
Is it important to you that you work in a \jpq{highly motivated small team}{SI11}{T015}?
With these job postings, you get to decide.
The \textit{Team Composition} sub-theme captures evidence describing the specific team you will work with.
This sub-theme has four codes: \textit{Diverse Team Composition}, \textit{Mono-Role Team Composition}, \textit{Intra-Team Benefit}, and \textit{Small Team}.
This sub-theme has a smaller amount of evidence compared to others, but can still be found in 22 of the 100 \gls{se} job postings.
In addition to the examples above, evidence for this sub-theme includes benefits from the team composition, such as \jpq{you will be part of a project-oriented team with very competent colleagues}{GL11}{T028}.

\subsubsection{Collaboration Skills}

Many job postings are looking for a \jpq{team player}{IN06}{T033} who can \jpq{work in a team collaborative environment}{SI01}{T016}.
If you are \jpq{comfortable with speaking at all organizational levels}{SI09}{T052} and have a \jpq{demonstrated ability to synthesize and present information}{MO19}{T016}, then you are the right candidate for these job postings.
The \textit{Collaboration Skills} sub-theme covers evidence discussing the candidate's ability to work with others, including their communication, collaboration, and networking skills.
This sub-theme is highly prevalent, with evidence in 68 of the 100 postings.
It includes six codes: \textit{Teamwork Ability}, four types of collaboration skills (\textit{Communication}, \textit{Interpersonal}, \textit{Presentation}, and \textit{Networking}), and \textit{Language Requirements}.
Networking skills, for example, are reflected in postings that emphasize the \jpq{ability to effectively establish and maintain working relationships and gather project requirements from stakeholders and clients}{SI02}{T013}.

\subsubsection{Self Management}

\gls{se} companies seek candidates who take initiative and demonstrate accountability in their work.
The \textit{Self Management} sub-theme highlights the importance of independence, responsibility, and initiative.
It has six codes, including personal characteristics such as \textit{Initiative}, \textit{Independence \& Autonomy}, and \textit{Responsibility \& Ownership}, and process characteristics including \textit{Follow Through on Commitments}, \textit{Time Management}, and \textit{Quality Work through Details and Rigour}.
Postings emphasize the need for a \jpq{self-starter, with a growth mindset}{IN02}{T022} and a \jpq{strong sense of ownership and accountability}{LI10}{T025}.
Candidates are expected to \jpq{have autonomy to research and achieve outcomes}{SI08}{T023} while also having \jpq{strong time management}{LI15}{T019}.
Postings also stress the importance of delivering high-quality work, seeking individuals who are \jpq{rigorous and concerned by the quality of the output delivered}{MO04}{T027} and maintain a \jpq{high level of attention to detail}{GL17}{T035}.

\subsubsection{Discussion}

Collaboration is a core expectation for software engineers, with evidence appearing in 90\% of job postings.
Notably, the code \textit{Communication Skills} has the most supporting evidence across all identified codes, highlighting that employers seek candidates who can clearly convey ideas, engage with stakeholders, and navigate team dynamics.
This signals the need for universities to strengthen communication training in \gls{se} curricula, particularly for cross-disciplinary teamwork.

Beyond communication, postings emphasize structured collaboration processes, including agile workflows, cross-functional interaction, and even non-hierarchical environments.
Self-management is also framed as a collaborative skill, reinforcing that autonomy and accountability are key to teamwork.
These findings suggest that, in the view of software companies, successful collaboration not only relies on interpersonal skills, but also on adaptability and personal responsibility.

\resultsSectionHeader{When and Where}{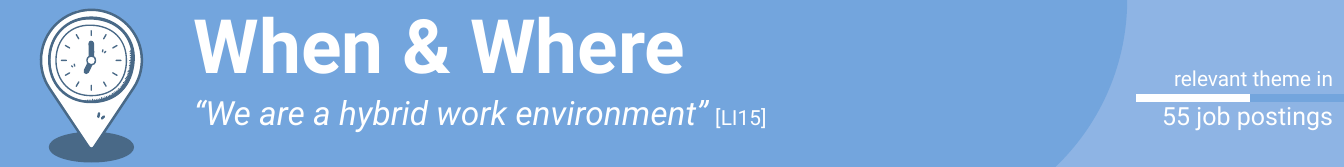}{fig:header_when_where}{Banner image indicating the start of the When and Where theme section. It features an icon representing the theme, the theme name in bold text, and the representative quote 'We are a hybrid work environment' from job posting LI15. It indicates that 55 of the 100 analysed job postings include evidence related to this theme.}

\emph{When \& Where} explores the expectations around working location, incentives for office presence, and time-off policies. Our analysis identified supporting evidence in 55 of the 100 job postings, across three sub-themes and 14 codes.

\subsubsection{Working Location}
Very few job postings in our sample specify explicit on-site attendance requirements (e.g., \jpq{Please only apply if you are able to reliably commute to the office 5 days a week}{LI13}{T019}), likely because this is still assumed to be the default in many cases. Statements regarding the possibility or even expectation of \jpq{100\% remote work}{IN14}{T030} or \jpq{100\% Home Office}{GL13}{T008} appear somewhat more frequently.

By far, the most evidence within this sub-theme supports flexible location choices: \jpq{Choose between our impressive office in Chancery Lane, remote work or a bit of both}{LI11}{T053}, and \jpq{That means if you want to do your thing in the office (if you're near one), at home or a bit of both, it's up to you}{SI06}{T010}.
Some postings specify a hybrid mix of remote and on-site work:
\jpq{The role is preferably hybrid, with 2 days per week spent in office}{LI14}{T025};
\jpq{While we can accommodate a position that is primarily remote this position does require the ability to work on-site as needed for meetings and team collaboration in our Norwich, CT office}{LI15}{T005}; and
\jpq{Christchurch based with hybrid working, our connection days in the office enable our supportive, collaborative, friendly and fun culture}{IN10}{T004}.
Others offer different work arrangements within their organization but make them dependent on the specific role or individual: \jpq{Click here to learn about our work personas: flexible, remote and required-in-office}{MO10}{T036}, \jpq{Onsite Flexibility: Preferred on-site in Redmond, WA but open to remote for strong candidates}{MO18}{T005}.

\subsubsection{Incentive to Work from Office}
Many \gls{se} job postings recognize the value of flexible working locations for employees, yet we also found evidence of efforts to make on-site work more appealing.
Some companies highlight the workplace itself as an incentive, promoting features such as a \jpq{Workplace Nursery Benefit}{LI11}{T094}, or a \jpq{friendly working environment located in a renovated farm in the countryside}{MO02}{T027}.
Others emphasize high-end office spaces, describing them as \jpq{fancy office with beautiful \& fancy pantry/meeting room/working office}{GL03}{T079}, a \jpq{modern and innovative working environment}{GL20}{T043}, or \jpq{an attractive workplace with modern equipment}{GL08}{T034}.

Beyond office amenities, some job postings promote the broader location as a benefit. Examples include offices \jpq{on the Dorset coast with its bustling quay, some of the best beaches in the UK and a stunning natural harbour}{LI13}{T015}, \jpq{in Innsbruck, in the middle of the Alps}{SI16}{T026}, or \jpq{in sunny Lisbon}{GL10}{T002}. To ease commuting, a few companies offer support, such as \jpq{free parking space, relieving the burden of parking expenses and streamlining your daily commute}{MO11}{T021} and, for example, a \jpq{1-year ticket for public transport [...] or a parking space near the office}{SI16}{T028}. In rare cases, companies even provide relocation assistance to help candidates move closer to the workplace {\scriptsize(e.g.~[SI13], [MO16])}.

\subsubsection{Time Off}

Building on the previously discussed flexibility in workplace location, many job postings also emphasize adaptable time-off policies. Some highlight particularly generous leave allowances, such as \jpq{up to 35 days off per year}{MO02}{T028}, while others offer incremental benefits over time, like sabbaticals after three and five years~{\scriptsize[LI11]} or \jpq{an additional day off granted every 5 years, allowing you to take a longer break, recharge, and cherish personal time for enhanced well-being and overall satisfaction}{MO11}{T024}.

Flexibility extends beyond the number of days off.
Some companies \jpq{offer the flexibility to take your local country's bank holiday allowance for other religious or cultural days}{LI11}{T069}, and offer \jpq{new parents a competitive matched parental leave as well as a phased return to work from extended leave}{LI11}{T093}.
Additional leave options include time for personal well-being or volunteering, such as \jpq{additional paid time to volunteer and give back to the community}{SI09}{T117}.
Many postings also highlight \jpq{flexible working times}{GL02}{T023} and \jpq{flexible working hours}{SI07}{T049}, sometimes explicitly linked to \jpq{wellness and wellbeing}{SI04}{T051} and the ability \jpq{to manage personal and professional responsibilities while at the same time ensur[ing] business operational needs and customer service expectations are achieved}{SI04}{T058}.

\subsubsection{Discussion}

Since we sampled job postings from 2023, some trends were likely still influenced by the lingering effects of COVID-19, which had normalized flexibility in work location~\cite{Ralph:2020:Pandemic}.
While some companies embraced remote work as an ongoing opportunity, others appeared to be shifting back toward on-site expectations.
Particularly in postings promoting hybrid work, the messaging suggests an attempt to combine the perceived advantages of both home office and in-person collaboration.

At the same time, job postings do not merely frame office presence as a requirement, but also highlight incentives to make on-site work more attractive---ranging from modern office environments to perks like workplace nurseries or desirable office locations.
This raises interesting questions for future research: How do such incentives influence developers' preferences and productivity? What factors make hybrid arrangements most effective? Understanding how companies balance flexibility with structured office presence could help shape best practices for supporting software engineers in diverse work settings.

\resultsSectionHeader{Explicit Perks}{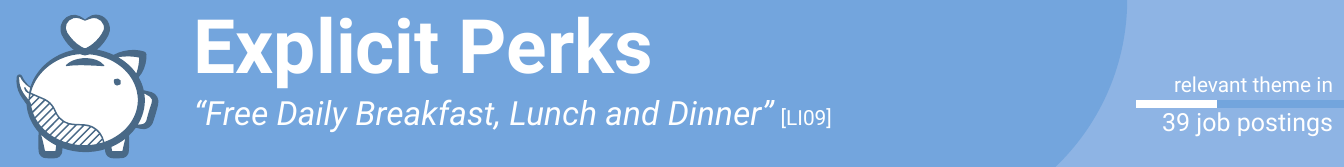}{fig:header_perks}{Banner image indicating the start of the Explicit Perks theme section. It features an icon representing the theme, the theme name in bold text, and the representative quote 'Free Daily Breakfast, Lunch and Dinner' from job posting LI09. It indicates that 39 of the 100 analysed job postings include evidence related to this theme.}
\label{sec:explicit_perks}

Job postings often include references to explicit perks---benefits that extend beyond role-specific aspects and highlight broader advantages of working for the company. During our thematic coding, we identified these perks as a distinct theme, independent of, for example, company culture, or job-specific expectations. While they are often listed toward the end of postings, their presence was not limited to dedicated benefits sections. This theme consists of three sub-themes: \emph{Remuneration}, \emph{Physical and Mental Health Offers}, and \emph{Insurance Offers}, comprising a total of 14 codes.

\subsubsection{Remuneration}

As a reminder, all job postings in our sample are for paid full-time positions. 
Nevertheless, some job postings make a point of highlighting their \jpq{attractive and competitive Salary}{GL03}{T080}, with successful candidates even being promised \jpq{a total compensation package that ranks among the best in the industry}{SI01}{T043}.
Specific figures are rarely mentioned when referring to competitive salaries, though there are exceptions, such as a Senior Software Engineer position in Portugal: \jpq{Competitive Salary: total expected compensation range between €55.000 and €85.000 / year}{GL10}{T012}.
This is likely because the actual offer depends heavily on the individual, as suggested by evidence pieces stating that salaries vary based on experience and position: \jpq{Competitive salary, depending on experience}{MO16}{T008}, \jpq{You will be offered an attractive compensation package based on your qualifications and the demands of this position}{GL11}{T029}.

In addition to a base salary, some positions offer an \jpq{annual performance bonus}{LI09}{T018}, and less frequently, stock options---either as a \jpq{Stock Option Plan}{LI07}{T034} or the opportunity for employees to purchase company stock: \jpq{Employees are also able to purchase company stock through our Employee Stock Purchase Program}{SI09}{T118}.

\subsubsection{Physical and Mental Health Offers}

The profession of a software engineer often involves spending long hours seated in front of a computer. Some companies aim to provide physical balance, for example, through \jpq{access to an on-site gym, encouraging a healthy lifestyle and fostering a positive work-life balance}{MO11}{T020}, \jpq{Health and Wellness focus including virtual fitness classes, wellness webinars, and discounted gym memberships}{IN07}{T033}, or a \jpq{cycle to work scheme}{MO15}{T044}.

Few workplaces go beyond fitness offerings to promote well-being by also providing \jpq{the choice of a wellbeing subsidy}{LI11}{T080}, a \jpq{company wellbeing programme}{SI19}{T045}, or \jpq{additional wellbeing leave}{LI20}{T007}.
At one company, \jpq{all employees also get access to a free counselling support service}{LI11}{T082}, while another offers \jpq{an annual Vibe \& Thrive allowance to support your wellbeing, social connection, office setup \& more}{SI06}{T040}.
A job posting for a Senior Software Engineer in Uruguay is more specific: \jpq{Our benefits show that we care about the whole you, from adoption and surrogacy assistance to tuition reimbursement and wellness programs}{GL04}{T006}.

Food and drink offerings are surprisingly rarely mentioned in our sample of job postings for software engineers. These include \jpq{Free Daily Breakfast, Lunch and Dinner}{LI09}{T022}, \jpq{Allowance / vouchers for food}{GL02}{T027}, or listings such as: \jpq{Free coffee/milk/detox drink/smoothies/juice/beer/wine/fruit/snack}{GL03}{T078}.

\subsubsection{Insurance Offers}

Insurance-related perks are explicitly mentioned in 20 job postings, typically in a straightforward manner, listing offerings such as \jpq{Health Care Plan (Medical, Dental \& Vision)}{LI07}{T030} or \jpq{Premium Medical, Dental and Vision Plan (No cost from employees or their families)}{LI09}{T021}.
Some postings highlight additional benefits, such as \jpq{psychological health}{SI04}{T047} coverage, \jpq{life and income protection}{IN10}{T041}, or employer-matched retirement plans, as seen in one posting stating that \jpq{U.S. employees have access to [\ldots] a 401(k) plan with a [employer] matching contribution}{SI09}{T114}.
Notably, health insurance appears to be omitted in postings from countries where it is a legal standard rather than a competitive benefit, suggesting that companies primarily emphasize perks that differentiate them in their specific labour markets.

\subsubsection{Discussion}

Many of the explicit perks highlighted in job postings---such as health insurance, gym memberships, or employer-matched retirement plans---are not necessarily unique to \gls{se} roles.
These benefits are commonly found across various industries, making them somewhat generic.
However, their inclusion in job postings suggests that they still play an important role in attracting talent.
\gls{se} roles are often distinguished by their focus on innovation and problem-solving, but similar opportunities may be available at multiple companies. In these cases, explicit perks can become a deciding factor.

A well-structured benefits package, including wellness or mental health programs, can signal an employer's commitment to well-being beyond core job duties.
However, it remains unclear whether these perks primarily serve to boost productivity (e.g., fitness programs to reduce sick days, mental health services to ensure focus) or reflect a genuine investment in long-term employee satisfaction~\cite{Graziotin:2019:Happiness}.
As job postings rarely elaborate on this distinction, future research could explore how employees perceive and experience these benefits in practice.

Some perks may also reveal employer expectations.
For instance, performance-based bonuses and stock options suggest an emphasis on long-term commitment and contribution to company success, while wellness programs and mental health support could imply an expectation of high-performance environments where resilience and stress management are crucial.
Future research could investigate whether and how candidates interpret these perks as signals of workplace culture and job demands.

\subsection{Cross-Theme Relations and Peculiarities}
\label{sec:cross-theme}

To further understand the relationships among the non-technical themes, we performed a frequent-set analysis based on the presence or absence of each theme in individual job postings.
In this context, a \emph{set} is defined by the combination of themes present.
For instance, 17 postings include evidence for all themes, while seven mention only Culture and Collaboration.
Notably, the top three sets account for 51\% of the postings.
This indicates a strong conformity to specific content patterns, with many postings consistently including evidence for the first four major themes while occasionally omitting aspects such as \emph{When and Where} and \emph{Explicit Perks}.
Furthermore, only three postings do not mention any non-technical aspects at all, focusing solely on technical and educational qualifications.
While we do not assert that non-technical aspects are more or less prevalent than technical ones, these results confirm that non-technical aspects are a meaningful part of modern \gls{se} hiring practices, reinforcing our initial motivation for this study.

Storytelling in job postings is an important part of convincing people to apply, since applicants can otherwise just use the meta-data searching features of job platforms to show them the exact jobs they qualify for.
Employers want to convince people to apply, and therefore how job postings are structured contributes to their success.
To investigate this storytelling aspect, we conducted a positional analysis to determine where themes appear within job postings.
Using violin plots, we visualized the relative position of each evidence piece within the text, revealing distinct placement patterns.
Themes such as \emph{Sense of Purpose} typically occur early (median 27\% of the way into the job posting), whereas \emph{Explicit Perks} tend to appear at the end (median 85\%).
Some sub-themes deviate from their parent themes, such as \emph{Specific Interests} appearing later despite belonging to \emph{Sense of Purpose}. 
Within \emph{Collaboration}, the sub-themes \emph{Collaboration Processes} and \emph{Team Composition} are noted earlier than \emph{Collaboration Skills} and \emph{Self Management}.
\emph{DEI Compliance} and \emph{DEI Appreciation} are consistently found at the end, suggesting they function as concluding or obligatory elements.
These positional trends highlight implicit structuring within job postings, which could be explored further to understand whether theme placement impacts candidate perception or application rates.
For additional details, further figures, scripts, and results are provided in our replication package~\cite{ReplicationPackage}.

\section{Implications and Broader Reflections}

Our analysis surfaces a wide range of non-technical aspects communicated through SE job postings.
These findings hold relevance for multiple stakeholders.

\paragraph{Implications for educators and curriculum design.}
Prior work has stressed the importance of integrating interpersonal skills like communication and teamwork into SE education~\cite{Garcia:2025:NonTechSkillsIndustryGap}.
Our findings reinforce this and extend the focus to additional non-technical aspects valued in practice but often overlooked in educational discourse, such as sense of purpose, personal growth, and company culture.
We observed notable variation in how companies address these themes: some offer structured support, others rely on individual initiative, or omit such expectations altogether.
This variation offers a productive starting point for critical reflection in SE education---e.g., in ethics or professional development courses---on what students may expect from employers and what responsibility they themselves carry in shaping their work environment and individual development.

\paragraph{Implications for software engineers and jobseekers.}
Our thematic map can support behavioural interview preparation and career planning by helping candidates reflect on what matters to them and whether a given job posting signals alignment.
For instance, if personal growth is a priority, candidates can note whether development opportunities are mentioned---and if not, feel confident raising the issue. 
Since many employers address such topics explicitly, their absence may itself be meaningful.
More broadly, our findings encourage jobseekers to assess employers more deliberately, treating postings not just as a list of requirements, but as curated representations of organizational values.

\paragraph{On the evolving role of software engineers.}
While tracing long-term changes would require longitudinal analysis, our study offers a detailed snapshot of how SE roles are currently framed: as collaborative, communicative, and embedded in sociotechnical systems.
The thematic map captures the breadth and nuance of these portrayals and contributes a data-driven perspective to ongoing conversations about how the profession is evolving.
As noted earlier, our analysis reflects a specific moment in time, but offers a valuable puzzle piece toward understanding broader shifts in workplace expectations.

\paragraph{On companies' self-presentation and the reality of work.}
We acknowledge that job postings may not perfectly reflect actual workplace practices or may omit unfavourable job-related information~\cite{Phillips:1998:RealisticJob}.
They may express employer intentions or serve branding purposes rather than document day-to-day realities~\cite{Jain:2015:Employment,Hein:2025:Branding}.
Our study does not assess whether these stated expectations are fulfilled.
Rather, it focuses on how roles and values are presented to jobseekers.
As curated communication artifacts, postings reveal what companies choose to emphasize in the hiring process.
The thematic map thus offers a valuable lens on this normative landscape---regardless of whether signals align with organizational reality.
Future work could investigate this potential gap more directly, e.g., by triangulating employer statements with employee experiences.

\section{Related Work}

Prior studies have analysed SE job postings using various methodologies, including keyword searches, topic modelling, and manual annotation.
For instance, \citet{Ehlers:2015:Socialness} examined job postings from Indeed and StackOverflow Careers by using a keyword-based search to identify common co-occurrences with the adjective \emph{social} in its immediate context. They found that companies frequently highlight social aspects such as team events, flexible work, or health care perks. 
From a similarly quantitative perspective, \citet{Montandon:2021:SkillsSOpostings} studied job postings by creating a word cloud of the 100 most frequent soft skill terms in 315 Stack Overflow ads.
They found \emph{communication} and \emph{collaboration} to play a central role.
This is in line with another job posting study by \citet{Rabelo:2022:NonTechSkillsSE}, who came up with a list of 30 \emph{non-technical skills} from 113 Stack Overflow ads, derived by manual annotation but without forming themes.
This is also consistent with a study by \citet{Gurcan:2019:ExpertiseRequiredSE} who used tokenization and computational topic modelling to identify \emph{communication} as a key theme.
All of these findings further align with our own impressions, particularly in the emphasis on teamwork and communication.
However, the referenced studies limit their exploration explicitly to the mention of \emph{skills} and, while useful for term spotting, the rather quantitative methodology of these studies misses semantically related content and higher-level thematic structures. 
Our \gls{ta} complements this work by surfacing implicit or under-explored expectations through a robust qualitative lens.

A study similar to our qualitative approach is presented by \citet{Galster:2023:SoftSkillsSE} in an analysis of job postings from New Zealand. 
They focused on inductively coding \emph{soft skills}, found such skills in 82\% of all postings, and related them to company size and core business. As the authors concluded after a subsequent tool-based soft skill extraction---conducted to compare with their manual findings---such a context-based investigation is currently only reliable through manual analysis.
Again, various communication skills were the most frequently demanded soft skills.
Similar regionally scoped studies, such as those conducted in Finland~\cite{Niva:2023:JuniorSECommCollab} and Ecuador~\cite{Chevez:2023:SkillsEcuador}, further reinforce the relevance of non-technical expectations like communication, collaboration, self-management, and creativity. 

\citet{Florea:2018:SoftwareTester} analysed postings for software \emph{testers} and found that 64\% mentioned soft skills, with, again, communication being the most common. 
The authors used an existing coding scheme from a previous study, which we deliberately chose not to do in our own analysis. Existing soft skill frameworks from SE or HR, while valuable, would not have captured the diverse and sometimes subtle themes that emerged in our data.
Our study followed an inductive, reflexive approach to Thematic Analysis, aiming to explore the full breadth of non-technical expectations in software job postings---not limited to soft skills alone.
Apart from that, the study by \citeauthor{Florea:2018:SoftwareTester} is interesting, as the requirements for software testers appear to differ little from those for software engineers, as examined in our study.

\citet{Rabelo:2022:NonTechSkillsSE} found in interviews with 15 junior developers that \emph{willingness to learn} is considered among the top non-technical skills required in their organizations, and five of the interviewees believe non-technical skills to not only play out on the individual level but to also influence team performance.

Despite differences in scope and method, these studies consistently reinforce our findings and complement our \gls{ta}.
The main difference lies in that related work specifically examines the expectations employers have for potential candidates' skills, whereas our study explores the broader sociotechnical context communicated in job postings.
This is ultimately reflected in that the findings of all listed studies each represent only a subset of our thematic map.

Beyond job postings, \gls{se} hiring also involves how criteria are communicated and perceived. An analysis of 10,000 Glassdoor reviews found that unclear hiring expectations often led candidates to perceive the process as arbitrary and unfair~\cite{Behroozi:2020:DebuggingHiring}. Similarly, companies sometimes overemphasize trending technologies in job ads---a practice termed \enquote{Résumé-Driven Development}---potentially misrepresenting actual job requirements~\cite{Fritzsch:2021:RDD, Fritzsch:2023:RDD}.

Recruiters' perspectives offer further insights into hiring challenges. A focus group study with SE recruiters identified common hiring antipatterns and provided recommendations to help candidates navigate these challenges~\cite{Setubal:2024:SERecruitment}. Technical interviews, for instance, have been found to assess not only a candidate's technical proficiency but also their ability to communicate and engage in problem-solving discussions~\cite{Ford:2017:TechTalk}. Candidate assessments may also be influenced by individual characteristics; for example, research on coding challenge performance found that sadness and high conscientiousness correlated with lower scores~\cite{Wyrich:2019:CodingChallenges}.

Finally, studies on software engineers' success emphasize adaptability, curiosity, and collaboration. Interviews with Microsoft engineers identified these as key attributes of a great software engineer, alongside technical expertise~\cite{Li:2015:WhatMakesAGreatSE}. Similarly, surveys with industry professionals highlight communication, teamwork, and leadership as essential non-technical skills~\cite{Garcia:2025:NonTechSkillsIndustryGap}. 
On a concluding note, the existence of many practitioner-oriented resources, such as books and blog posts on \emph{Building Great Software Engineering Teams}~\cite{Tyler:2015:BuildingGreatSETeams} and \emph{Cracking the Coding Interview}~\cite{McDowell:2015:Cracking}, reflect a broader industry interest in effective hiring and candidate preparation.
SE research echoes this interest through a renewed focus on interview preparation in academic contexts~\cite{Kapoor:2025:InterviewPrep,Bell:2025:InterviewPrep}

\section{Conclusion}

The evolving narrative of what makes a great software engineer extends beyond technical expertise, embracing a broader set of qualities reflected in job postings. Our thematic analysis provides a snapshot of these non-technical aspects, offering a structured look at how employers present roles, company culture, and candidate expectations.
While postings sometimes present a curated view of these aspects, they do not necessarily capture the full reality of workplace experiences---a gap that future research could explore by comparing advertised expectations with employee experiences. 
As an overview study, our work lays the groundwork for additional investigations into each of the identified themes, supporting a more holistic understanding of Software Engineering as not just a technical discipline, but a sociotechnical one.

\subsubsection*{{\normalfont\textbf{Data Availability}}}

Our comprehensive replication package~\cite{ReplicationPackage} ensures full transparency and reproducibility, including job posting data, analysis scripts, decision protocols, and evidence tracing.
It also features an interactive visualization of the thematic map, allowing users to explore the data themselves.

\bibliographystyle{ACM-Reference-Format}
\bibliography{main}

\end{document}